\input phyzzx
\overfullrule=0pt\hsize=6.5truein\baselineskip=24pt\voffset=0pt\vsize=8.9 in


\def\br{\hfil\break} \def\rarr{\rightarrow}
\def\scrscr{\scriptscriptstyle}  
\def\spil{height2pt&\omit&&\omit&&\omit&&\omit&\cr}
\def\ar{\cr\spil\noalign{\hrule}\spil} \def\LA{\Lambda}
\def\dsp{\displaystyle} \def\w#1{\;\hbox{#1}\;}
\def\X#1{_{\lower2pt\hbox{$\scrscr#1$}}} \def\lL{\left|}\def\rL{\right|}
\def\ns#1{_{\hbox{\sevenrm #1}}}
\def\w#1{\;\hbox{#1}\;}\def\const{\hbox{const.}}
\def\ph{\phi}\def\pt{\partial}\def\OM{\Omega}\def\dd{{\rm d}}
\def\scr{\scriptstyle}\def\Z#1{_{\lower2pt\hbox{$\scr#1$}}}
\def\g#1{{\rm g}\Z#1}\def\RN{Reissner-Nordstr\"om}
\def\SI{\Sigma} \def\G#1{{\rm g}\X#1} \def\V{{\cal V}} 
\def\pZ#1{\ph\Z{#1}}  
\def\sgn{\w{sgn}} \def\rH{r\X{\cal H}}\def\RH{R\X{\cal H}} \def\e{{\rm e}}
\def\pH{\ph\X{\cal H}} \def\GG{\G0^{\ 2}} \def\th{\theta} \def\PH{\Phi}
\def\al{\alpha}\def\OO#1{{\rm O}\left(#1\right)}\def\la{\lambda}
\def\Ve{\V\ns{exp}}\def\Vs{\V\ns{susy}}\def\ds{{\rm ds}^2}\def\kp{2\phi}

\def\dVdph{{\dd\V\over\dd\ph}}\def\lb{\bar\la}
 \def\gbb{\left(\g1\pm\g0\right)} 
\def\GG{\G0^{\ 2}} \def\GGG{\G1^{\ 2}} 
\def\be{\beta} \def\thm#1#2{\noindent{\bf #1}\quad{\it #2}} \def\rh{\rho}
\def\Proof{\par\noindent{\bf Proof}\quad} \def\epph{\e^{2\G0\ph\X0}}
\def\Qph{Q^2\e^{2\G0\ph}} \def\Pph{P^2\e^{-2\G0\ph}} \def\ta{\tau}
\def\hp#1{\hat\ph\Z#1} \def\hf#1{\hat f\Z#1} \def\hR#1{\hat R\Z#1}

\ifx\jams\undefined  
\def\FOOT#1{\foot{#1}}
\def\ABSTRACT{\baselineskip=18pt\abstract}
\def\endpage{\centerline{(December, 1994)}\vfill
\centerline{Published in {\it J.\ Australian Math.\ Soc.}\ {\bf B41} (1999)
198-216.}\vfill\baselineskip=16pt plus.2pt minus.1pt\vfil\eject}
\def\begincaption{\medskip\openup-1\jot\begingroup\Tenpoint}
\def\endcaption{\endgroup\openup1\jot\leftskip=0pt\rightskip=0pt}
\def\caption#1#2{\message{#1}\begincaption\leftskip=15true mm\rightskip=15true
mm\vbox{\halign{\vtop{\parindent=0pt\parskip=0pt\strut##\strut}\cr{\bf#1}\quad
#2\cr}}\endcaption}
\def\inserttable{\midinsert \input\jobname.tab \endinsert}
\def\tableout{}
\rightline{\vbox{\halign{#\hfil\cr ADP-94-29/M26\cr gr-qc/9502038\cr}}}
\bigskip\bigskip
\else 
\vsize=9.5true in\def\FOOT#1{} 
\def\ABSTRACT{\centerline{\bf Abstract} \baselineskip=20pt plus.2ptminus.1pt}
\def\endpage{\centerline{{\bf Short title}: Dilaton black holes}\vfill\eject}
\def\caption#1#2{\noindent{\bf #1} #2}
\def\inserttable{\midinsert\centerline{TABLE 1 NEAR HERE}\endinsert}
\chapterskip=120pt \def\tableout{\vfil\eject \input\jobname.tab}
\fi
\sectionstyle{\bf}\def\section#1{\par \ifnum\lastpenalty=30000\else\penalty-200
\vskip\sectionskip\spacecheck\sectionminspace\fi\global\advance\sectionnumber
by 1 {\protect\xdef\sectionlabel{\the\sectionstyle{\the\sectionnumber}} \wlog
{\string\section\space\sectionlabel}}\noindent{\caps\enspace\sectionlabel.~~#1}
\par\nobreak\vskip\headskip \penalty 30000 }
\def\appendix#1{\par \ifnum\lastpenalty=30000\else\penalty-200
\vskip\sectionskip\spacecheck\sectionminspace\fi\global\usechapterlabeltrue
\chapterreset \xdef\chapterlabel{#1} \wlog{\string\Appendix~\chapterlabel}
\noindent{\caps\enspace Appendix}\par\nobreak\vskip\headskip \penalty 30000}
\def\refout{\par\penalty-400\vskip\chapterskip\spacecheck\referenceminspace
\ifreferenceopen \Closeout\referencewrite \referenceopenfalse \fi
\line{\bf References\hfil}\vskip\headskip\input\jobname.refs}

\title{\seventeenbf Dilaton Black Holes\break With a Cosmological Term\FOOT
{Based on a talk given at the {\it Inaugural Australian General
Relativity Workshop}, Australian National University, Canberra, September,
1994; to appear with the proceedings in {\it J.\ Australian Math.\ Soc.\
{\bf B}}}}
\vfill

\author{David L. Wiltshire\FOOT{E-mail: dlw@physics.adelaide.edu.au}}

\address{Department of Physics and Mathematical Physics, University of
Adelaide,\break Adelaide, S.A. 5005, Australia.} \vfill

\ABSTRACT
The properties of static spherically symmetric black holes, which carry
electric and magnetic charges, and which are coupled to the dilaton in
the presence of a cosmological constant, $\LA$, are reviewed.
\vfill
\endpage
 
\def\PL#1{{\it Phys.\ Lett.}\ {\bf#1}}
\def\PRL#1{{\it Phys.\ Rev.\ Lett.}\ {\bf#1}}
\def\AP#1#2{{\it Ann.\ Phys.}\ (#1) {\bf#2}}
\def\PR#1{{\it Phys.\ Rev.}\ {\bf#1}}
\def\CQG#1{{\it Class.\ Quantum Grav.}\ {\bf#1}}
\def\NP#1{{\it Nucl.\ Phys.}\ {\bf#1}}
\def\GRG#1{{\it Gen.\ Relativ.\ Grav.}\ {\bf#1}}

\REF\AGM{P.S. Aspinwall, B.R. Greene and D.R. Morrison, ``Calabi-Yau moduli
space, mirror manifolds and space-time topology change in string theory'',
\NP{B416} (1994) 414--480.}
\REF\CHM{K.C.K. Chan, J.H. Horne and  R.B. Mann, ``Charged dilaton black holes
with unusual asymptotics'', \NP{B447} (1995) 441--464.}
\REF\Cho{Y.M. Cho, ``Reinterpretation of Jordan-Brans-Dicke theory and
Kaluza-Klein cosmology'', \PRL{68} (1992) 3133--3136.}
\REF\DP{T. Damour and A.M. Polyakov, ``The string dilaton and a least coupling
principle'', \NP{B423} (1994) 532--558.}
\REF\DIN{J.P. Derendinger, L.E. Ib\'a\~nez and H.P. Nilles, ``On the low energy
$d=4$, $N=1$ supergravity theory extracted from the $d=10$, $N=1$
superstring'', \PL{155B} (1985) 65--70.} 
\REF\DRSW{M. Dine, R. Rohm, N. Seiberg and E. Witten, ``Gluino condensation in
superstring models'', \PL{156B} (1985) 55--60.}
\REF\DM{P. Dobiasch and D. Maison, ``Stationary, spherically symmetric
solutions of Jordan's unified theory of gravity and electromagnetism'',
\GRG{14} (1982) 231--242.}
\REF\faya{H.F. Dowker, J.P. Gauntlett, D.A. Kastor and J. Traschen, ``Pair
creation of dilaton black holes'', \PR{D49} (1994) 2909--2917.}
\REF\fayb{H.F. Dowker, J.P. Gauntlett, S.B. Giddings and G.T. Horowitz,
``On pair creation of extremal black holes and Kaluza-Klein monopoles'',
\PR{D50} (1994) 2662--2679.} 
\REF\GHS{D. Garfinkle, G.T. Horowitz and A. Strominger, ``Charged black holes
in string theory'', \PR{D43} (1991) 3140--3143; (E) {\bf45} (1992) 3888.}
\REF\Gi{G.W. Gibbons, ``Antigravitating black hole solitons with scalar hair in
$N=4$ supergravity'', \NP{B207} (1982) 337--349.}
\REF\GHT{G.W. Gibbons, G.T. Horowitz and P.K. Townsend, ``Higher-dimensional
resolution of dilatonic black hole singularities'', \CQG{12} (1995) 297--317.}
\REF\GK{G.W. Gibbons and R.E. Kallosh, ``Topology, entropy and Witten index of
dilaton black holes'', \PR{D51} (1995) 2839--2862.}
\REF\GM{G.W. Gibbons and K. Maeda, ``Black holes and membranes in higher
dimensional theories with dilaton fields'', \NP{B298} (1988) 741--775.}
\REF\GW{G.W. Gibbons and D.L. Wiltshire, ``Black holes in Kaluza-Klein
theory'', \AP{N.Y.}{167} (1986) 201--223; (E) {\bf176} (1987) 393.}
\REF\GSW{M.B. Green, J.H. Schwarz and E. Witten, {\it Superstring theory},
(Cambridge University Press, 1987).}
\REF\GMS{B.R. Greene, D.R. Morrison and A. Strominger, ``Black hole
condensation and the unification of string vacua'', \NP{B451} (1995) 109--120.}
\REF\GH{R. Gregory and J.A. Harvey, ``Black holes with a massive dilaton'',
\PR{D47} (1993) 2411--2422.}
\REF\GP{D.J. Gross and M.J. Perry, ``Magnetic monopoles in Kaluza-Klein
theories'', \NP{B226} (1983) 29--48.}
\REF\HHR{S.W. Hawking, G.T. Horowitz and S.F Ross, ``Entropy, area and black
hole pairs'', \PR{D51} (1995) 4302--4314.}
\REF\HW{C.F.E. Holzhey and F. Wilczek, ``Black holes as elementary particles'',
\NP{B380} (1992) 447--477.}
\REF\HH{J.H. Horne and G.T. Horowitz, ``Black holes coupled to a massive
dilaton'', \NP{B399} (1993) 169--196.}
\REF\HoHo{J.H. Horne and G.T. Horowitz, ``Cosmic censorship and the dilaton'',
\PR{D48} (1993) R5457--R5462.}
\REF\KT{D. Kastor and J. Traschen, ``Cosmological multi-black hole solutions'',
\PR{D47} (1993) 5370--5375.}
\REF\MaSo{G. Magnano and L.M. Soko{\l}owski, ``On physical equivalence between
nonlinear gravity theories and a general-relativistic scalar field'', \PR{D50}
(1994) 5039--5059.}
\REF\MaSh{T. Maki and K. Shiraishi, ``Multi black hole solutions in
cosmological Einstein-Maxwell-dilaton theory'', \CQG{10} (1993) 2171--2178.}
\REF\MWa{S. Mignemi and D.L. Wiltshire, ``Spherically symmetric solutions in
dimensionally reduced spacetimes'', \CQG{6} (1989) 987--1002.} 
\REF\MWb{S. Mignemi and D.L. Wiltshire, ``Black holes in higher derivative
gravity theories'', \PR{D46} (1992) 1475--1506.}
\REF\Ok{T. Okai, ``4-dimensional dilaton black holes with cosmological
constant'', Preprint UT-679, hep-th/9406126 (1994).}
\REF\PW{S.J. Poletti and D.L. Wiltshire, ``Global properties of static
spherically symmetric spacetimes with a Liouville potential'', \PR{D50} (1994)
7260--7270; (E) {\bf D52} (1995) 3753--3754.}
\REF\PTW{S.J. Poletti, J. Twamley and D.L. Wiltshire, ``Charged dilaton black
holes with a cosmological constant'', \PR{D51} (1995) 5720--5724; (E) {\bf D52}
(1995) 3753-4.} 
\REF\ptw{S.J. Poletti, J. Twamley and D.L. Wiltshire, ``Dyonic dilaton black
holes'', \CQG{12} (1995) 1753--1769, 2355.}
\REF\Ros{S.F. Ross, ``Pair production of black holes in a $U(1)\times U(1)$
theory'', \PR{D49} (1994) 6599--6605.}
\REF\Soka{L.M. Soko{\l}owski, ``Uniqueness of the metric line element in
dimensionally reduced theories'', \CQG{6} (1989) 59--76.} 
\REF\Sokb{L.M. Soko{\l}owski, ``Physical versions of non-linear gravity
theories and positivity of energy'', \CQG{6} (1989) 2045--50.}
\REF\Sor{R.D. Sorkin, ``Kaluza-Klein monopole'', \PRL{51} (1983) 87--90; (E)
{\bf54} (1985) 86.}
\REF\Str{A. Strominger, ``Massless black holes and conifolds in string
theory'', \NP{B451} (1995) 96--108.}
\REF\tHa{G. 't Hooft, ``Strings from gravity'', {\it Phys.\ Scr.}\ {\bf T15}
(1987) 143-150.}
\REF\tHb{G. 't Hooft, ``The black hole interpretation of string theory'',
\NP{B335} (1990) 138--54.}
\REF\Ts{A. Tseytlin, ``Exact solutions of closed string theory'', \CQG{12}
(1995) 2365--2410.}
\REF\Wil{D.L. Wiltshire, ``Spherically symmetric solutions in dimensionally
reduced spacetimes with a higher-dimensional cosmological constant'', \PR{D44}
(1991) 1100--1114.}

\footsymbolcount=0\def\foot{\advance\footsymbolcount by 1
\attach{\the\footsymbolcount}\Vfootnote{$^{\the\footsymbolcount}$}}

\immediate\openout\figurewrite=\jobname.tab \begingroup
\catcode`|=0\catcode`\$=12\catcode`\#=12\catcode`\&=12\catcode`\\ =12
|immediate|write|figurewrite{\centerline{\vbox{
\offinterlineskip\hrule\halign{&\vrule#&\strut\quad$\dsp#$
\quad\hfil\cr\spil&\omit&&\lL f\rL&&\lL h\rL&&e^{2\ph}&\ar&K\Z{1,2}&&R^{2/\GG}
&&\const&&R^{\pm2/\G0}&\ar&M\Z{1,2}&&\const&&\const&&\const&\ar&N\Z{1,2}&&R^2&
&R^{2(1-\GGG)}&&R^{2\G1}&\ar&P\Z{1,2}&&R^{2/\GGG}&&\const&&R^{2/\G1}&\ar
&T\Z{1,2}&&R^{2\be\X1}&&R^{2\be\X2}&&R^{4/(\G1\pm\G0)}&\cr\spil}\hrule}}}
|immediate|write|figurewrite{\caption{Table 1}
{Asymptotic form of singly charged solutions for trajectories
approaching critical points at phase space infinity from within the sphere at
infinity, in the case that $R\rarr\infty$. The points are labelled as in [\PW]
-- in some cases the values of $\g0$, $\g1$, $\lb$ and $\LA$ are restricted if
certain points are to exist.\br Here $\be\Z1=1+2\left[2-\g1\gbb\right]/\gbb^2$
and $\be\Z2=2\g1/\gbb-1$.\br In all instances the $+$ sign refers to electric
solutions, and the $-$ sign to magnetic solutions.}}
|endgroup
\closeout\figurewrite
\section{Introduction}

String theory [\GSW] appears to provide the most promising framework we know
of for constructing a quantum theory of gravity, a goal which has remained
elusive despite the attentions of at least one generation of researchers.
Unfortunately, although the mathematical structure of string theory is very
rich, it has proved to be extremely difficult to extract a phenomenologically
viable model of physics from this structure. In particular, in compactifying
the extra spatial dimensions of the fundamental superstring theory on a
Calabi-Yau manifold, one finds a plethora of possible 4-dimensional vacua, with
no means of determining just which one should correspond to the groundstate of
the universe. However, recent work suggests that this central problem of
string phenomenology might be resolved through the introduction of a single
large ``universal'' moduli space [\AGM], whose various branches correspond to
distinct Calabi-Yau vacua. To consistently deal with conifold singularities
which arise in this moduli space, it is necessary to include certain black
hole states in the spectrum of string excitations [\GMS], which have the
unusual property of being massless [\Str]. Thus there appears to be an intimate
connection between string theory and black holes. Such a relationship has been
suggested before on various grounds [\tHa,\tHb], and over the last five years a
number of exact black hole solutions of string theory have
been found [\Ts]. There would thus appear to be much insight to be gained from
studying processes involving such stringy black hole states. At a more modest
level, we might hope to gain some understanding of stringy black holes by
studying the properties of solutions to the effective field theoretic models
which arise in the low-energy limit of string theory -- namely, the models one
obtains when one keeps only the lowest-order terms in a perturbative expansion
in powers of the Regge slope parameter, $\al'=1/(2\pi T)$, $T$ being the
string tension. Indeed a number of such solutions can be promoted to be
solutions of the exact string theory to all orders in $\al'$ [\Ts], which
justifies this approach.

In this paper I would like to discuss recent work (with S. Poletti and J.
Twamley) [\PW--\ptw] concerning the properties of black holes coupled to the
dilaton field in the presence of a cosmological term. The scalar dilaton, along
with the pseudoscalar axion, enters into the action of effective gravitational
theories that arise in the low-energy limit of string theory. Although one can
consistently truncate such models to consider, for example, solutions with a
vanishing axion, the presence of the dilaton cannot be ignored if one is to
consider the coupling of gravity to other gauge fields, and thus the dilaton
can be considered to be an essential feature of stringy gravity.

The effective string action we will consider takes the form
$$S=\int\dd^4x\sqrt{-\hat g}\,e^{-2\ph}\left\{{\hat{\cal R}\over4}+\hat g^{\mu
\nu}\pt_\mu\ph\,\pt_\nu\ph-\hat\V(\ph)-{1\over4}F_{\mu\nu}F^{\mu\nu}+\dots
\right\}\eqn\saction$$
in four dimensions\foot{I will work exclusively in four dimensions here. The
extension to arbitrary $D\ge4$ is straightforward (at least as far as the
non-numerical results are concerned), and is dealt with in [\PW--\ptw].}, where
we have included the dilaton $\ph$, with potential $\V(\ph )$, and an abelian
$U(1)$ gauge field with field strength $F_{\mu\nu}$ as representative of matter
content. The axion, which would enter with the matter degrees of freedom in a
similar fashion to $\bf F$, has been excluded. It is a characteristic of string
theory that the dilaton, $\ph$, couples universally to matter through the
conformal factor $e^{-2\ph}$ in \saction. The metric $\hat g_{\mu\nu}$ is thus
said to correspond to the {\it string conformal frame}. To obtain an action
resembling that of general relativity we conformally rescale to obtain the
metric
$$g_{\mu\nu}=\e^{-2\ph}\hat g_{\mu\nu}\eqn\contran$$
of the {\it Einstein conformal frame}, in which the action becomes
$$S=\int\dd^4x\sqrt{-g}\left\{{{\cal R}\over4}-{1\over2}\,g^{\mu\nu}\pt_\mu\ph
\,\pt_\nu\ph-\V(\ph)-{1\over4}\e^{-2\G0\ph}F_{\mu\nu}F^{\mu\nu}+\dots\right\},
\eqn\action$$
where $\V\equiv\e^{2\ph}\hat\V$, and $\g0=1$ in the case of string theory. It
is the non-trivial coupling between $\ph$ and $\bf F$ in \action\ which leads
to a theory with rather different properties than that of the standard
Einstein-Maxwell ($\g0=0$) theory.

Although the value of the dilaton coupling parameter is fixed to be $\g0=1$ by
the tree-level string action, it is nonetheless useful to leave this as an
arbitrary parameter since a variety of similar theories can then be treated
together. Another case of interest is $\g0=\sqrt{3}$, which corresponds to the
4-dimensional theory obtained by dimensionally reducing the 5-dimensional
Kaluza-Klein theory. In that case the scalar $\ph$ is associated with the
radius of the fifth dimension rather than being a dilaton field. From the
string theoretic point of view the potential $\V(\ph)$ can be expected to be
present if one is dealing, for example, with a central charge deficit in which
case the potential takes the Liouville form
$$\Ve={\LA\over2}\,\e^{-2\G1\ph},\eqn\Vexp$$
with $\g1=-1$ and $\LA=\left(D_{\hbox{\sevenrm crit}}-D_{\hbox{\sevenrm eff}}
\right)/(3\al')$, $\al'$ being the Regge slope parameter. Alternatively, if one
considers non-perturbative effects such as the generation of a dilaton mass by
supersymmetry breaking then one might also expect a contribution to the
potential of the form
$$\Vs=\exp\left[-\al e^{-\kp}\right]\left\{A\Z1e^{\kp}+A\Z2+A\Z3e^{-\kp}\right
\}\,.\eqn\Vsusy$$
where $\al$, $A\Z1$, $A\Z2$ and $A\Z3$ are constants. This is in fact the
potential which arises in four dimensions from supersymmetry breaking via one
gaugino condensation in the hidden sector of the string theory [\DIN,\DRSW],
and should be regarded as typical of the general type of potential one might
expect.

There has been a great deal of interest in the past few years in black hole
solutions of effective low-energy theories of gravity derived from string
theory, such as those described above. Most interest has focussed in particular
on black holes with a massless dilaton, $\V\equiv0$. For $\g0=1$, which is the
case of interest to string theory the general solution is given by\foot{The
radial coordinate of Gibbons and Maeda [\GM] is obtained by $r\rarr r+M$, and
that of Garfinkle {\it et al}.\ [\GHS] by $r\rarr r-\SI$, relative to the
coordinate chosen here. Our choice here is dictated by the requirement that $R
(r)$ agrees with the gauge chosen later in asymptotic expansions.}
[\GHS,\Gi,\GM]
$$\eqalignno{\ds&=-f\dd t^2+f^{-1}\dd r^2+R^2\dd\OM^2,&\eqname\coorda\cr{\bf F}
&={Q\over R^2}\,\e^{2\ph}\dd t\wedge\dd r+P\sin\th\Z1\dd\th\Z1\wedge\dd\th\Z2,&
\eqname\FF\cr\e^{2\ph}&={r+\SI\over r-\SI}&\eqname\sisi\cr}$$
where
$$\eqalignno{R(r)\equiv&\,\sqrt{(r+\SI)(r-\SI)},&\eqname\RR\cr f(r)\equiv&\,{1
\over R^2}\left[(r-M)^2-(M^2+\SI^2-Q^2-P^2)\right],&\eqname\ff\cr2M\SI=&\,P^2-Q
^2,&\eqname\conA\cr}$$
and $\dd\OM^2$ is the standard round metric on a $2$-sphere, with angular
coordinates $\th\Z1$, $\th\Z2$. The scalar charge $\SI$ has been normalised
such that
$$\phi={\SI\over r}+\OO{r^{-2}}\eqn\assphi$$
at spatial infinity. It is possible to add an arbitrary constant $\ph\Z0$ to $
\ph$ at the expense of normalising the electromagnetic field differently from
the standard choice at spatial infinity; however, we shall not do so here.
Similar exact solutions are known for $\g0=\sqrt{3}$ in the case of dyonic
solutions [\DM], and for all values of $\g0$ in the case of solutions with a
single (electric or magnetic) charge [\GM].

Unlike the standard \RN\ solutions the global properties of the solutions vary
considerably according to whether the solutions carry a single charge, or both
electric and magnetic charges. In the case of the latter dyonic solutions the
spacetime exhibits two horizons and the properties of the solutions are similar
to those of the \RN\ solutions. In particular, the extremal limit, $|P|+|Q|=
\sqrt{2}\,M$, [or $\SI=(|P|-|Q|)/\sqrt{2}\;$], corresponds to the two horizons
becoming degenerate at a finite value of $R$, and the black holes have zero
Hawking temperature in this limit. However, the singly charged solutions have a
single horizon, $\rH$, and the extreme solutions are obtained in the limit $\RH
\equiv R(\rH)\rarr0$, and $\pH\equiv\ph(\rH)\rarr\infty$ for magnetic solutions
($Q=0$), (or $\pH\rarr-\infty$ for electric solutions ($P=0$)). As a result the
area of the event horizon and the entropy of the extreme solutions is zero.
Furthermore, the extreme solutions have finite Hawking temperature if $\g0=1$,
and infinite temperature if $\g0>1$ [\GM,\GW]. The fact that the temperature is
formally infinite of course merely signals the breakdown of the semi-classical
approximation if one is considering the Hawking evaporation process. Holzhey
and Wilczek [\HW] have in fact demonstrated that in the case of the $\g0>1$
solutions an infinite mass gap develops for quanta with a mass less than that
of the black hole so that the Hawking radiation slows down and comes to an end
at the extremal limit, despite the infinite temperature. Furthermore, since
scalar waves with an energy below the mass of the black hole are repelled the
black holes behave in a fashion similar to that of elementary particles [\HW].

An interesting feature of this extreme limit is that in the case of the purely
magnetic solutions the spacetime is in fact completely regular when viewed in
terms of the metric $\hat g_{\mu\nu}$ of the string conformal frame [\GHS].
Since strings couple to $\hat g_{\mu\nu}$ string physics is not affected by the
singularity which appears at $R=0$ in the Einstein conformal frame in the
extreme case. In the string conformal frame the $R=0$ singularity is replaced
by an infinitely long throat [\GHS]. The property that the singularity here
resides entirely in the conformal factor is also a feature of other solutions.
In particular, in the case $\g0=\sqrt{3}$ the purely magnetic solution
corresponds to the Kaluza-Klein monopole [\GP,\Sor], which is singular from the
point of view of the 4-dimensional Einstein conformal frame but is completely
regular in five dimensions. Further examples are discussed in [\GHT]. Although
arguments based on the positivity of energy favour the Einstein conformal frame
as being the physical one in the context of Jordan-Brans-Dicke theories and
higher derivative models [\Cho,\MaSo,\Soka,\Sokb], it is clear that
singularities which can be removed by a conformal rescaling -- so that roughly
speaking the singularity is in the Ricci tensor rather than the Weyl tensor --
may be milder than other types of singularities. However, some work remains to
be done to put this on a firmer footing in some general framework if one does
not wish to appeal to string theory or higher-dimensional physics.

The fact that classical solutions are obtained in dilaton gravity with
properties that differ significantly from the standard Einstein-Maxwell theory
has provided fertile ground for the development of new ideas concerning various
quantum gravitational phenomena, such as pair creation
[\faya,\fayb,\GK,\HHR,\Ros] and fission
[\GK] of black holes. These ideas have led to the realisation that the entropy
of extreme black holes should be zero, even in the case of the \RN\ solutions
which has a horizon of finite area in the extreme limit [\HHR].

One major defect with the dilaton black hole models studied to date is that the
contribution of possible dilaton terms, $\V(\ph)$, has been largely ignored.
Phenomenologically this could be regarded as a major defect, since it is widely
believed that the dilaton must acquire a mass in order to avoid the generation
of long range scalar forces, which are not observed in nature. This is by no
means the only possibility -- Damour and Polyakov have recently suggested an
interesting scenario in which the dilaton remains massless but very weakly
coupled to ordinary matter [\DP] -- however, the ``massive dilaton'' scenario
remains the conventional choice, and certainly one that deserves to be further
studied. The reason why such models have not been studied very extensively is
simply that the problem of finding appropriate exact solutions is a technically
very difficult one. Gregory and Harvey [\GH] and Horne and Horowitz [\HH] have
made an investigation of black hole models which include a mass term, in the
form of a simple quadratic potential [\GH,\HH],
$\V=2m^2\left(\ph-\ph\Z0\right)^2$, or alternatively of the form [\GH]
$\V=2m^2\cosh^2\left(\phi-\ph\Z0\right)$. While a rigorous proof of the
existence of black hole solutions in these models has still to be given, the
arguments of Horne and Horowitz [\HH] based on numerical work are nonetheless
compelling. The stability of the solutions remains an open question, however.
Given that the putative black hole solution of [\HH] appears to be the envelope
of a set of singular solutions for $R(r)$ and a separatrix between different
singular solutions for $\ph(r)$, it seems doubtful that this solution could be
stable -- although the situation may be different for other potentials.
However, it is clear from the arguments of [\GH,\HH] that the properties of the
solutions with a massive dilaton are essentially the same as those with a
massless dilaton in the case of black holes which are small with respect to the
Compton wavelength of the dilaton. This is a feature which one might expect to
be generic of any potential, and it does provide some further justification for
studying the solutions with a massless dilaton.

The aim of our recent series of papers [\PW--\ptw] has been to study a problem
which is similar to that of dilaton black holes with a mass-generating
potential, but is technically somewhat simpler. In particular, we have set out
to derive the properties of solutions with a cosmological term, which might be
expected to be analogues of the \RN-(anti)-de Sitter solutions. I will outline
our major results in turn.

\section{Static spherically symmetric solutions with a Liouville potential}

If one is to consider a cosmological term in the context of stringy gravity
then the first question to be asked is what is the appropriate cosmological
term to be considered. Rather than taking a simple cosmological constant, a
Liouville-type potential \Vexp\ may be more appropriate, as one can thereby
obtain the potential appropriate to a central charge deficit. Indeed,
Kastor-Traschen type [\KT] cosmological multi-black hole solutions have been
discussed in the context of such models by Horne and Horowitz [\HoHo], and by
Maki and Shiraishi [\MaSh]. However, all exact solutions which have been
constructed with non-zero dilaton couplings obtained involve a dilaton which
depends on certain special powers of the time-dependent cosmic scale factor
[\MaSh], and thus in particular they possess no static limit.

On the face of it, it would appear to be simple to rule out physically relevant
black hole solutions in models with a potential \Vexp, on the grounds that they
do not possess ``realistic'' asymptotics. In particular, if by ``realistic''
asymptotics we require that black holes be asymptotically flat or
asymptotically of constant curvature we might conjecture that:

\noindent (i)\ If $\V$ is non-zero then asymptotically (anti)-de Sitter
solutions exist if and only if
$$\exists\;\ph\Z0\ \w{such that}\ \left.\dVdph\right|_{\ph=\ph\X0}=0\ \w{and}\
\V(\ph\Z0)\ne0.\eqn\conddS$$
The solutions are asymptotically de Sitter (anti-de Sitter) for $\V(\ph\Z0)>0$
($\V(\ph\Z0)<0$).

\noindent (ii)\ If $\V$ is non-zero then asymptotically flat solutions exist if
and only if
$$\exists\;\ph\Z0\ \w{such that}\ \left.\dVdph\right|_{\ph=\ph\X0}=0\ \w{and}\
\V(\ph\Z0)=0.\eqn\condaf$$
The fact that trivial solutions exist under both these circumstances is quite
obvious: if \conddS\ holds then the Schwarzschild-(anti)-de Sitter solutions
with constant dilaton, $\ph=\ph\Z0$, are solutions; while if \condaf\ is
satisfied then the Schwarzschild solution with constant dilaton, $\ph=\ph\Z0$,
is a solution. Any particular potential may have many such solutions, depending
on the number of different turning points.

The proof of necessity is much less clearly defined, since in the context of
stringy gravity it is not obvious what are physically realistic restrictions to
place on the dilaton asymptotically. The approach one would take in a
conventional field theory setting is to demand that all fields have regular
Taylor expansions at spatial infinity, which in terms of the coordinates
\coorda\ with $f=f(r)$, $R=R(r)$, are given by
$$\eqalign{\ph=\;&\ph\Z0+{\ph\Z1\over r}+{\ph\Z2\over r^2}+\dots,\cr f=\;&{-\LA
r^2\over3}+f\Z{-1}r+f\Z0+{f\Z1\over r}+{f\Z2\over r^2}+\dots,\cr R=\;&r+R\Z0+{R
\Z1\over r}+{R\Z2\over r^2}+\dots,\cr}\eqn\expand$$
if the solutions are assumed to be asymptotically flat or (anti)-de Sitter
depending on the value of $\LA$. If we now take the independent field equations
obtained by variation of the action \action, viz.\
$$\eqalignno{\left[R^2f\ph'\right]'=\;&{{\dd\V}\over{\dd\ph}}R^2+{\g0\over R^2}
\left[\Qph-\Pph\right]\,,&\eqname\feaA\cr{R''\over R}=\;&-\ph'^2,&\eqname\feaB
\cr\left[f\left(R^2\right)'\right]'=\;&2-4\V R^2-{2\over R^2}\left[\Qph+\Pph
\right]\,,&\eqname\feaC\cr}$$
where $'\equiv d/dr$, and substitute the expansions \expand, we obtain the
result
$$\left.\dVdph\right|_{\ph=\ph\X0}=0,\qquad\LA=2\V(\ph\Z0),\eqn\asympV$$
from the lowest order terms, which proves the conjecture. ``Realistic'' black
hole solutions are consequently ruled out for the Liouville-type potential
\Vexp, since it is monotonic in $\ph$, except in the special case of a
cosmological constant ($\g1=0$) when $\dVdph\equiv0$ identically.

In the context of stringy gravity, however, the asymptotic condition on $\ph$
given in \expand\ is too restrictive. In particular, in string theory all
couplings between the dilaton and matter fields involve powers of $e^{2\ph}$,
so that provided the dilaton energy-momentum tensor is well-behaved at spatial
infinity one would expect the {\it weak-coupling limit} $\ph\rarr-\infty$ to be
physically admissable asymptotically. In [\PW] we have therefore given a proof
of necessity of the above conjecture in the case of singly charged black hole
solutions with a Liouville-type potential \Vexp, without making any
restrictions on the asymptotic behaviour of $\ph$. Our approach generalises a
similar method used to derive the global properties of uncharged static
spherically symmetric solutions in a variety of models involving scalar fields
with potentials [\MWa,\Wil,\MWb]. Unfortunately, we have been unable to extend
the method
to deal with arbitrary potentials, $\V(\ph)$, and so are unable to offer a
proof of the conjecture except on a case by case basis.

The argument of [\PW] proceeds by using an alternative radial coordinate [\GM],
defined by $\dd\xi=f^{-1}R^{-2}\dd r$, and a slightly generalised metric,
namely
$$\dd s^2=f\left[-\dd t^2+R^{4}\dd\xi^2\right]+R^2g_{ij}\dd x^i\dd x^j,\eqn
\coordb$$
where now $f=f(\xi)$, $R=R(\xi)$, and $g_{ij}$ is a 2-dimensional metric of
constant curvature
$${\cal R}_{ij}=\lb\,g_{ij},\qquad i=1,2.\eqn\emetric$$
Spherically symmetric solutions have $\lb>0$ -- however, integral curves in the
$\lb>0$ portion of the phase space can also have endpoints with $\lb=0$, which
is why this generalisation is necessary. It is then possible to convert the
field equations into a 5-dimensional autonomous system of ordinary differential
equations, and consequently various properties of the solutions can be derived
from the 5-dimensional phase space using standard techniques from the theory of
dynamical systems. The crucial results which concern us are:\br (i)\ Critical
points at finite values of the phase space parameters consist of a 2-parameter
set with $\lb=0$, $\LA=0$ and $Q=0$ (or $P=0$ as appropriate). Solutions with $
\lb>0$ with endpoints at such critical points are found to correspond either to
central singularities, or in the case of a 1-parameter subset, to horizons.\br
(ii)\ Critical points at phase space infinity can be divided up according to
whether $R\rarr0$ (central singularity), $R\rarr\const$ (Robinson-Bertotti
solutions), or $R\rarr\infty$ (asymptotic region).

The asymptotic properties of the solutions are conveniently summarised in terms
of the metric functions one obtains when using $R$ as the radial variable,
viz.\
$$\dd s^2=-f\dd t^2+h^{-1}\dd R^2+R^2\dd\OM^2\eqn\coordb$$
where $h(R)\equiv f\left(\dd R\over\dd r\right)^2$. The possible asymptotic
behaviour of the solutions is summarised in Table 1.\inserttable
The points $M\Z{1,2}$ correspond to the asymptotically flat solutions which lie
in the subspace with a vanishing potential, $\V(\ph)\equiv0$, and which of
course include the black hole solutions of [\GM,\GHS]. The only other instance
in which ``realistic'' asymptotics are obtained is the special case of a
cosmological constant, $\g1=0$, when points $N\Z{1,2}$ correspond to
asymptotically de Sitter or anti-de Sitter regions, according to the sign of
$\LA$. The weak coupling limit for the dilaton is attained in various cases for
points other than $M\Z{1,2}$, however, the metric is not asymptotically flat or
of constant curvature in any of these instances.

One must add the caveat that the points $K\Z{1,2}$ provide an interesting
special case in which solutions exist with a Riemann tensor whose components
in an orthonormal frame vanish as $1/R^2$ at spatial infinity, even though the
asymptotic structure of the spatial hypersurfaces at spatial infinity is not
that of flat space [\PW]. Solutions
with endpoints at $K\Z{1,2}$ will include a 1-parameter subset with regular
horizons [\CHM,\PW], and in certain cases the dilaton will be asymptotically
weakly coupled. Chan, Horne and Mann [\CHM], who have studied these exotic
black hole solutions in detail, argue that such solutions could be
physical. Clearly, the physical properties of solutions with such unusual
asymptotics deserve further study. However, we will put aside these interesting
questions and will make the more conservative choice of demanding that the
solutions be asymptotically flat or asymptotically (anti)-de Sitter. With this
proviso the only possibile ``realistic'' black hole solutions with a
non-trivial dilaton potential would be those corresponding to integral curves
from outside the $\V\equiv0$ subspace which nonetheless have endpoints at
$M\Z{1,2}$. An analysis of small perturbations about $M\Z{1,2}$ reveals that
there are no solutions in this category, however. This
completes the proof of the non-existence of charged black hole solutions for
the Liouville potential \Vexp\ in all cases other the pure cosmological
constant, $\g1=0$.

\section{Static spherically symmetric solutions with a cosmological constant}

In the case of a cosmological constant an analysis of small perturbations
around $N\Z{1,2}$ reveals that such points attract or repel a 5-dimensional set
of trajectories. To further show that black hole solutions exist it is
necessary to show that there exist integral curves linking points $N\Z{1,2}$ to
the 1-parameter set of critical points mentioned above which correspond to
regular horizons. This seems reasonable given that the dimensionality of the
set of solutions with endpoints at $N\Z{1,2}$ is greater than that of the other
critical points. However, although it might be possible to give a rigorous
argument simply from the properties of the phase space, the large
dimensionality of the phase space makes it difficult. There is one further
complication if $\LA>0$ since in that case the asymptotic de Sitter region
corresponds to one in which the Killing vector $\pt/\pt t$ is spacelike. The
outermost horizon is therefore a cosmological one, and to obtain genuine black
hole solutions at least two horizons are required. In terms of the phase space
this means an additional requirement of the existence of integral curves which
link different points in the 1-parameter family of solutions that correspond to
horizons. Again the large dimensionality of the phase space makes the
resolution of this question difficult. Instead we have found that the issue can
be settled by a simple argument relating to the original dilaton field equation
\feaA.

\thm{Theorem 3.1}{The maximum number of regular horizons possessed by a static
spherically symmetric spacetime coupled to the dilaton in the presence of a
cosmological constant, $\V\equiv\LA/2$, and an electromagnetic field which is
either electrically charged ($Q\ne0$, $P=0$) or magnetically charged ($P\ne0$,
$Q=0$) with arbitrary coupling constant, $\g0$, is one\foot{\rm For $\LA>0$
this result is more restrictive than that obtained by Okai [\Ok], who also
placed limits on the number of possible horizons by a different approach.}.}
\advance\footsymbolcount by 1
\Proof A straightforward proof by contradiction was given in [\PTW]. Consider
the case $P=0$ and $\g0>0$. Suppose that the spacetime possesses at least two
regular horizons, and let the two outermost horizons be labelled $r_\pm$, with
$r_-<r_+$. Regularity implies that near $r=r_+$, $f\propto(r-r_+)$ and $\ph(r_
+)$ and $R(r_+)$ are bounded with $R(r_+)\ne0$, and similarly for $r_-$. From
\feaA\ it follows that at both horizons
$$\ph'f'\,\Bigr|_{r_\pm}={\g0\Qph\over R^2}\,\Biggr|_{r_\pm}\eqn\horz$$
Suppose that $\pt/\pt t$ is timelike in the asymptotic region, as is
appropriate for asymptotically flat or asymptotically anti-de Sitter solutions,
so that $f(r)<0$ on the interval $(r_-,r_+)$, $f'(r_-)<0$ and $f'(r_+)>0$. For
such solutions \horz\ implies that $\ph'(r_-)<0$ and $\ph'(r_+)>0$. These two
values of $\ph'$ must be smoothly connected and thus $\ph'(r)$ must go through
zero at least once in the interval $(r_-,r_+)$ at a point $r\Z0$ such that
$\ph''(r\Z0)>0$. However, since $f(r)<0$ on the interval $(r_-,r_+)$ it follows
from \feaA\ that if $\ph'(r\Z0)=0$ then $\sgn\ph''(r\Z0)=-\sgn\g0<0$. We thus
obtain a contradiction.

If $\pt/\pt t$ is spacelike in the asymptotic region, as is appropriate for
asymptotically de Sitter solutions, then $f'(r_-)>0$ and $f'(r_+)<0$, and thus
the signs of $\ph'(r_-)$ and $\ph'(r_+)$ are reversed in the above argument.
However, since $f(r)>0$ now on the interval $(r_-,r_+)$, the sign of $\ph''(r\Z
0)$ must also be reversed and one still obtains a contradiction. Similar
remarks apply if one takes the purely magnetic case ($Q=0$), or if one reverses
the sign of $\g0$, which completes the proof for all cases.

\thm{Corollary 3.2}{No static spherically symmetric asymptotically de Sitter
black hole solutions exist when coupled to the dilaton in the presence of a
cosmological constant, $\V\equiv\LA/2$, and an electromagnetic field which is
either electrically charged ($Q\ne0$, $P=0$) or magnetically charged ($P\ne0$,
$Q=0$) with arbitrary coupling constant, $\g0$.} \Proof As was noted above,
black hole solutions with $\LA>0$ must have at least two horizons, and so the
result follows.

It is clear that the above results will not apply to the case of dyonic
solutions, since then the r.h.s.\ of \feaA\ does not have a definite sign,
which was a crucial requirement of the argument based on \horz. Nevertheless,
similar restrictions can be made.

\thm{Theorem 3.3}{The maximum number of regular horizons possessed by a static
spherically symmetric spacetime coupled to a non-constant dilaton field in the
presence of a cosmological constant, $\V\equiv\LA/2$, and an electromagnetic
field which is both electrically and magnetically charged, $Q\ne0$ and $P\ne0$,
with arbitrary coupling constant, $\g0$, is:\br (i) two, if $\pt/\pt t$ is
timelike in the asymptotic region;\br (ii) one, if $\pt/\pt t$ is spacelike in
the asymptotic region.} \Proof As was shown in [\ptw] a simple proof by
contradiction is obtained by assuming the existence of at least two regular
horizons, $r_-$ and $r_+$, as previously. It is convenient to define a rescaled
dilaton
$$\PH\equiv\ph+{1\over2\g0}\ln\left|Q\over P\right|.\eqn\scaledil$$
If the dilaton in \feaA\ is rescaled in this fashion, and the resulting
equation multiplied by $\PH$ we obtain
$$\left[R^2f\PH\PH'\right]'=R^2f\PH'^2+2|QP|\,{\g0\PH\sinh(2\g0\PH)\over R^2}\,
.\eqn\Horz$$
This equation can be integrated with respect to $r$ between the horizons, $r_-$
and $r_+$, with the result that the integral of the l.h.s.\ vanishes. If $f(r)
>0$ on the interval $(r_-,r_+)$ then the r.h.s.\ of \Horz\ is
positive-definite, and a contradiction is obtained in all cases except that of
a trivial constant dilaton, $\PH=0$, $\PH'=0$. A succession of three regular
horizons is thus ruled out in the case of non-trivial solutions since this
would require that either $f(r)>0$ between the first pair of horizons or else
$f(r)>0$ between the second pair. Furthermore, we can have $f(r)<0$ between the
outermost horizons only if $f(r)>0$ on the interval $(r_+,\infty)$, i.e., if $
\pt/\pt t$ is timelike in the asymptotic region, which is true for
asymptotically flat and asymptotically anti-de Sitter solutions but not for
asymptotically de Sitter solutions. The latter solutions can thus have at most
one regular horizon if the dilaton is non-trivial.

\thm{Corollary 3.4}{The unique static spherically symmetric asymptotically de
Sitter black hole spacetime coupled to the dilaton in the presence of a
cosmological constant, $\V\equiv\LA/2$, and an electromagnetic field which is
both electrically and magnetically charged, $Q\ne0$ and $P\ne0$, with arbitrary
coupling constant, $\g0$, is the \RN-de Sitter solution with constant dilaton:}
$$\eqalign{R(r)&=r,\cr f(r)&={-\LA r^2\over3}+1-{2M\over r}+{2|QP|\over r^2}\,,
\cr\epph&=\left|P\over Q\right|.\cr}\eqn\rnds$$
\Proof Since black hole solutions with $\LA>0$ must have at least two horizons
and $\pt/\pt t$ is spacelike in the asymptotic region, it follows from \Horz\
that $\PH=0$, $\PH'=0$ is the only possible dilaton solution. The solution
\rnds\ is then obtained by direct integration.\smallskip

If we require that $\ph\Z0=0$, so as to obtain the conventional normalisation
for the electromagnetic field at spatial infinity, we see that \rnds\ is in
fact the solution for equal electric and magnetic charges, $|Q|=|P|$. The exact
solution \rnds\ of course also applies to $\LA\le0$. In addition, there also
exist Robinson-Bertotti-type solutions of the form $\epph=|P|/|Q|$, $R=R\ns{ext
}$ and
$$f=\left[R\ns{ext}^{-2}-4\LA\right]+c\Z1r+c\Z2\eqn\rbA$$
where $c\Z1$ and $c\Z2$ are arbitrary constants, and the constant $R\ns{ext}$
is a solution of the quartic
$$-\LA R\ns{ext}^4+R\ns{ext}^2-2|QP|=0.\eqn\rbB$$

\section{Asymptotically anti-de Sitter black holes}

In view of the results outlined in the last section it is clear that in the
presence of a cosmological constant, $\LA$, charged or dyonic black holes with
a non-trivial dilaton will only exist if $\LA<0$. Unfortunately there is no
obvious way of determining an exact solution in closed form. In [\PTW,\ptw] we
have therefore resorted to numerical integration.

If we solve for the series expansions \expand\ from the field equations we find
$$\eqalign{\ph=\;&{\pZ3\over r^3}+\OO{1\over r^4},\cr f=\;&{-\LA r^2\over3}+1-{
2M\over r}+{Q^2+P^2\over r^2}+{\LA\pZ3^{\ 2}\over5r^4}+\OO{1\over r^5},\cr R=\;
&r-{3\pZ3^{\ 2}\over20r^5}+\OO{1\over r^6},\cr}\eqn\assia$$
where we have made the gauge choice $R\Z0=0$, $\pZ0=0$, which is the same
choice made in \coorda--\ff. The effect of the dilaton is therefore short
range, as opposed to the asymptotically flat case \assphi. The black holes with
a massive dilaton [\GH,\HH] similarly give rise to a short range force --
however, in that case $\ph=\pZ4r^{-4}+\OO{r^{-5}}$ asymptotically. Furthermore,
$\pZ4$ is fixed in terms of the other charges directly by the asymptotic series
solution in the solutions of [\GH,\HH], whereas here $\pZ3$ is only fixed once
we make the further demand that particular solutions with the asymptotic form
\assia\ correspond to black holes.

The appropriate constraint on $\pZ3$ is found by integrating the dilaton
equation \feaA\ between the outermost horizon, $\rH$, and spatial infinity, to
give
$$\pZ3={\g0\over\LA}\int_{\rH}^\infty{\dd r\over R^2}\,\left(\Qph-\Pph\right)\,
.\eqn\intrel$$
This result is a direct analogue of \conA\ -- it fixes $\pZ3$ in terms of the
other charges of the theory, so that $\pZ3$ is not an independent ``hair''.

For the purposes of numerical integration it is convenient to use $R$ as the
radial variable, and to work with the metric \coordb. One can then solve for $f
$ in terms of $h$ and $\ph$: $f=h\exp\left[2\int^R\dd\bar R\,\dot\ph^2\bar R
\right]$, where $.\equiv\dd/\dd R$. Only two independent field equations then
remain, viz.\
$$\eqalignno{-&R\dot h+1-h=\left(\LA+h\dot\ph^2\right)R^2+\left(\Qph+\Pph
\right){1\over R^2},&\eqname\febA\cr&Rh\ddot\ph+s(h,\ph,\dot\ph)=0,&\eqname
\febB\cr}$$
where
$$s(h,\ph,\dot\ph)\equiv\dot\ph\left[1+h-\LA R^2-\left(\Qph+\Pph\right){1\over
R^2}\right]-{\g0\over R^3}\left(\Qph-\Pph\right).\eqn\shoot$$

The numerical procedure we adopted was to treat the problem as an initial value
one, and to integrate outwards from a regular horizon, $\RH$, the initial data
being set a small distance away from $\RH$ using the solutions obtained by
substituting the series expansions
$$\eqalign{h=\sum_{i=1}^\infty\tilde h_i\left(R-\RH\right)^i,\cr\ph=\pH+\sum_{i
=1}^\infty\tilde\pZ i\left(R-\RH\right)^i.\cr}\eqn\horexp$$
in the field equations \febA, \febB. For the lowest order terms, for example,
$$\eqalign{\tilde h\Z1&={1\over\RH^{\ 3}}\left[\RH^{\ 2}-\LA\RH^{\ 4}-Q^2\e^{2
\G0\pH}-P^2\e^{-2\G0\pH}\right],\cr\tilde\pZ1&={\g0\over\tilde h\Z1\RH^{\ 4}}
\left(Q^2\e^{2\G0\pH}-P^2\e^{-2\G0\pH}\right).\cr}\eqn\horsol$$
As the higher order terms are very lengthy we will not list them here. An outer
horizon is obtained if we restrict to initial data such that $\tilde h\Z1>0$.
We then integrate out until the solutions match the appropriate asymptotic
series expressed in terms of $R$, rather than in terms of $r$ as in \assia.
This is possible for a range of initial values, $\RH$, $\pH$. Details of the
numerical procedures are discussed in [\PTW,\ptw], and numerous resulting plots
are presented there. Our results indicate that asymptotically anti-de Sitter
black holes with a non-trivial dilaton do exist, and with all possible numbers
of horizons which are not excluded by Theorems 3.1 and 3.3. I will not
reproduce all these results here but will outline the important features.

\subsection{4.1\ {\it Dyonic solutions}} There is an additional numerical
complication for dyonic solutions since it is possible, but not guaranteed, for
asymptotically anti-de Sitter solutions to exist with two horizons in this
case. For singly charged solutions on the other hand two horizons are ruled out
by Theorem 3.1. To deal with the issue of the number of horizons, we first
examined the case of the asymptotically flat solutions with arbitrary coupling
constant, $\g0$, since it appears that exact dyonic solutions are only known in
the cases $\g0=1$ [\Gi,\GHS,\GM] and $\g0=\sqrt{3}$ [\DM,\GW]. Although the
singular nature of the differential equations \febA, \febB\ in general prevents
one integrating through the horizons, it is possible to determine whether one
is dealing with a genuine second horizon or not by examining the quantity $s(h,
\ph,\dot\ph)$ defined by \shoot. At a genuine horizon $s$ should approach zero
as $h$ approaches zero. Interestingly enough, our numerical results for the
asymptotically flat case indicate that we appear to be dealing with a
non-linear eigenvalue problem. These results [\ptw] are consistent with the
existence of two horizons only if $\g0$ is the square root of a triangle
number:
$$\g0=0,1,\sqrt{3},\sqrt{6},\dots,\sqrt{n(n+1)/2},\dots\eqn\evalue$$

In the case of the asymptotically anti-de Sitter solutions a similar
qualitative behaviour is observed, with two horizons only existing for certain
critical values of $\g0$, which differ slightly from the values \evalue.
However, whereas the critical values of $\g0$ are independent of the initial
conditions in the asymptotically flat case, this is no longer true for
asymptotically anti-de Sitter solutions. Thus the critical values would appear
to involve some complicated relationship between $\g0$, $\LA$, $M$ and $|QP|$.

As far as the thermodynamics properties of the solutions are concerned, there
is no qualitative difference between the cases of dyonic black holes with one
or two horizons. In particular, the pattern of isotherms [\ptw] is
qualitatively the same in both cases, and the Hawking temperature of the
solutions, which is given in general by
$$T={1\over4\pi\RH^{\ 3}}\left[\RH^{\ 2}-\LA\RH^{\ 4}-Q^2\e^{2\G0\pH}-P^2\e^{-2
\G0\pH}\right],\eqn\tempt$$
is always zero in the extreme limit. This is related to the fact that the
extreme solution always occurs at a finite value of $R$ determined from \rbB:
$$R\ns{ext}={-1\over2\LA}\left[\sqrt{1-8\LA|QP|}\,-1\right].\eqn\rbC$$
The spacetime geometry of the extreme solutions approaches that of the
Robinson-Bertotti-type solutions \rbA, \rbB\ in the neighbourhood of the
degenerate horizon. Furthermore, the entropy -- at least, the entropy
na\"{\i}vely defined as one quarter the area of the event horizon -- is finite
in the extreme limit, similarly to the \RN\ and \RN-anti de Sitter solutions.

\subsection{4.2\ {\it Singly charged solutions}} The solutions with a single
electric or magnetic charge are found to be qualitatively different to the
dyonic solutions. This would appear to stem from the fact that there are no
Robinson-Bertotti-type solutions in this case. As in the case of the
asymptotically flat dilaton black holes there is a maximum of one horizon and
the extreme limit corresponds to $\RH\rarr0$ and $\pH\rarr\infty$ for magnetic
solutions, (or $\pH\rarr-\infty$ for electric solutions), independently of the
value of $\LA$, assuming $\LA<0$ and $\g0>0$. Consequently, from \tempt\ we see
that the extreme solutions will have a formally infinite Hawking temperature,
and zero entropy. This is indeed verified by the numerical solutions [\PTW]. Of
course, the infinite temperature is merely a signal of the breakdown of the
semi-classical approximation.

In the case of the $\g0=1$ magnetic solutions, one can verify that the ``throat
structure'' of the string conformal frame metric of the asymptotically flat
extreme solutions is preserved for the asymptotically anti-de Sitter black
holes. In particular, consider the independent field equations derived from the
action \saction, with $\hat\V=\LA\e^{-2\ph}/2$, as is appropriate for an
Einstein frame cosmological term. With a string conformal frame metric of the
form
$$\ds=-\hat f\dd\ta^2+\hat f^{-1}\dd\rh^2+\hat R^2\dd\OM^2,\eqn\coords$$
$\hat f=\hat f(\rh)$, $\hat R=\hat R(\rh)$, these are given by
$$\eqalignno{\left[\hat R^2\hat f\left(\e^{-2\ph}\right)'\right]'&=2\e^{-2\ph}
{P^2\over\hat R^2},&\eqname\fecA\cr{\hat R''\over\hat R}&=-\ph'^2,&\eqname\fecB
\cr\left[\hat f\left(\hat R^2\e^{-2\ph}\right)'\right]'&=2\e^{-2\ph}-2\LA\hat R
^2\e^{-4\ph}-2\e^{-2\ph}{P^2\over\hat R^2}\,,&\eqname\fecC\cr}$$
where now $'\equiv\dd/\dd\rh$. In the asymptotically flat case [\GHS] the
neighbourhood of the $\pH\rarr\infty$ singularity is a so-called ``linear
dilaton vacuum'', i.e., of the form $\ph=-\al\rh$ as $\rh\rarr\infty$, ($\al<0$
being a constant), with vanishing corrections of the form $\e^{\al\rh}$. We
thus look for similar solutions of the form
$$\eqalign{\e^{-2\ph}&=\hp1\e^{\al\rh}+\hp2\e^{2\al\rh}+\dots,\cr\hat f&=\hf0+
\hf1\e^{\al\rh}+\dots,\cr\hat R&=\hR0+\hR1\e^{\al\rh}+\dots.\cr}\eqn\throat$$
It is straightforward to verify from the field equations \fecA--\fecC\ that the
following leading order coefficients are unaffected\foot{By contrast, it is
possible that $\hf0$ is altered for certain non-trivial dilaton potentials
[\GH].} by the presence of $\LA$: $\hp1$ (free), $\hp2=-2\hp1\hR1/\hR0$, $\hR0^
{\ 2}=2P^2$, $\hf0=1/(2P^2\al^2)$. Thus the extremal solution does indeed
appear to be a direct analogue of the one of Garfinkle, Horowitz and Strominger
[\GHS]. The cosmological constant only affects the higher-order corrections:
$\hf1=-\LA\hp1/\al^2$, $\hR1=c\Z1-2\sqrt{2}\LA\al P^3\hp1\rh/3$, etc.

\section{Conclusion}

If we neglect the possibility of black holes with an exotic asymptotic
structure [\CHM], then no physically interesting black hole spacetimes exist if
the dilaton has a
Liouville-type potential, except in the case that this potential is a simple
cosmological constant in the Einstein conformal frame. Black hole solutions
with a non-trivial dilaton only exist for a negative cosmological constant.
These spacetimes are asymptotically anti-de Sitter with a short range
correction due to the dilaton. The horizon structure and thermodynamic
properties of the solutions are the same as corresponding asymptotically flat
solutions with a massless dilaton in both the dyonic and singly charged cases.

Although a pure cosmological constant may not be the most natural cosmological
term in the context of stringy gravity, we hope that the solutions we have
studied here may nontheless provide a useful approximation in some
circumstances. In particular, if we have a non-trivial dilaton potential and if
the minimum of that potential which corresponds to the groundstate of the
dilaton has a value which is not precisely zero, then the universe would
contain some (hopefully small) vacuum energy. If this vacuum energy is negative
then the black hole solutions studied here (with $\LA<0$) might provide a
useful approximation to the complete solutions, just as the solutions of
[\GH,\HH] are well approximated by the solutions with a massless dilaton in
certain regimes. If there is to be a non-zero vacuum energy then
phenomenologically a small positive vacuum energy is the one favoured by
current measurements of the Hubble constant. Although we find no black hole
solutions with a non-trivial dilaton for $\LA>0$, it is possible that this
conclusion could be altered in the presence of a non-trivial dilaton potential,
$\V(\ph)$. In particular, the dilaton equation \feaA, upon which Theorems 1 and
2 are based, acquires an additional $\dd\V\over\dd\ph$ term which could provide
an obstruction to these arguments for suitable potentials. Of course, for
suitably complex potentials there are many additional features which could
further complicate things.

\medskip\noindent{\bf Acknowledgement}\quad I would like to thank Steve Poletti
and Jason Twamley for many useful discussions.
\refout
\tableout
\bye